\documentclass[
reprint,
twocolumn,
a4paper,
secnumarabic,
nofootinbib,
 amsmath,amssymb,
 aps,
]{revtex4-1}

\usepackage{graphicx}
\usepackage{dcolumn}
\usepackage{bm}

\begin{document}

\title{Physics beyond Causality}%

\author{A.V.~Novikov-Borodin}%
\affiliation{Institute for Nuclear Research of RAS, 60-th October Anniversary prospect~7a, 117312 Moscow, Russia}
\begin{abstract} 
The representations of the world around in physics built with help of causality are analyzed and seems incomplete. The observer's causal representations form a closed logical system, i.e. the compact group related to cause-effect chains. The space-time representations are exactly the background of this closed system and the off-site phenomena exceeded the space-time continuum of the observer are investigated. Off-site phenomena occur responsible for `mysteries and paradoxes' in quantum physics and `dark substances' in modern cosmology. The existing paradigm of cognoscibility is reconsidered and specified. The theory of sets predicts an infinite number of levels of cognition, where the world around seems more and more disordered and chaotic. The possibilities of different levels of cognition are estimated from this point of view. Relativistic and quantum theories operate on different levels of cognition, so their unification in frames of rigorous logical theory seems quite doubtful. 
\begin{description}
\item[Keywords]
Space-time, General relativity, Quantum theories, Theory of cognition.
\item[PACS numbers]
04.20.Cv, 03.65.Bz, 12.90.+b.
\end{description}
\end{abstract}
\maketitle
%
%
%
%
\section*{\bf Introduction}

Whether space-times, different from the observer's one, can exist? How can we describe them and what physical situation it may concern? How could we observe the physical objects from different space-times, and do they have correspondences with known ones? These questions together with arising epistemological problems are being considered and discussed. 

\section{\bf Space-Time}
\label{sec:ST}

The observer perceives events in the world around through his frame of references $S$: $x\equiv (x^0{,}\ldots{,}x^{n-1}) $, which is a set of continuous variables or coordinates, i.e. some reference marks or reference events in space and time. Due to relativistic principle the space-time and matter are equivalently described from a class of \emph{inertial} frames $S_T $ related with each other by a group of transformations $T$. It is the Galileo group in the Newtonian-Galilean scheme (NG) and they are generalized by enlarging to the Lorentz group in Special Relativity (SR) (see, for example, L.~Ryder Ref.\cite{Ryd09}, 2009). 

The `origin' of inertiality is postulated. In practice, to distinguish the inertial frame from non-inertial one, it is proposed to use a test particle. However, all particles belong to investigated matter by default, so the test particle and the observer himself are also parts of investigated matter. Thus, there is `hidden' axiomatic interconnection in SR (and in NG) of the space-time $G_{SR}$ perceived by the observer through inertial frames $S_{T}$ with matter ${\cal M} $: $ {G}_{SR}\left(S_{T}\right) \dashleftarrow  {\cal M}$. The space-time representations occur \emph{logically close} for the observer: \emph{he tests his own space-time by particles of matter, to which he belongs himself}. Off-site matter, off-site space-times, such parallel worlds, even if exist, are undistinguished for the observer from nonexistent. In SR (and NG) it looks equivalent to the conclusion that `parallel worlds' do not exist at all. 

The `hidden' axiomatic interconnections between space-time and matter become explicit in General relativity (GR), because now the gravitating matter can bend the space-time topology and SR representations of a space-time are generalized to the GR space-time manifold: $ {G}_{GR}\left( S_{T}; {\cal M}\right) \rightleftharpoons {\cal M}$. The affine connectivity and metrics are usually introduced in all considered models of the space-time, so in classical and relativistic theories (NG, SR and GR) the space-time is represented as a continuum. 

As before, the observer in GR perceives his own space-time, his `ordinary' matter to which he belongs himself, and these representations seem also \emph{logically closed} for him, but now this closure is explicit and may be described mathematically: \emph{``The general theory of relativity generates conservation laws inside itself and not in the form of consequences of the field equations, but in the form of identities. There is the following. If you are considering the integral ${I}=\int {\cal R} d^4 x$, in which ${\cal R}$ is, of course, an invariant density, so the four identical ratios between Hamiltonian derivatives of ${\cal R}$ just follow from the only one fact of the general invariance of this integral and these ratios are of type of conservation laws''} (E.~Schr\"odinger Ref.\cite{Schr50}). The existence of one closed system leads to the existence of others and E.~Schr\"odinger as long ago as 1950 had noticed to this possibility: \emph{``These identities are not the only ones. Any scalar density generates some system of identities. \ldots If one has two different densities, it may seem that their conservation laws and field equations do not have any correspondences at all, as though stay off-site from each other''}. 

Off-site continuums would be again `undistinguished from nonexistent', if nobody proposes the way of their interaction with the observer's continuum. An explicit influence of matter onto the space-time topology in GR gives us this possibility. We will consider that not only `ordinary' matter from observer's continuum, but also `off-site' matters $\tilde {\cal M}$ from off-site continuums $\tilde G$ (below in text we will usually mark off-site parameters with tilde) may be involved in gravitational interactions in other continuums. Thus, continuums occur interconnected in the Global system (GS): 
\begin{eqnarray}
\left\{ \begin{array}{l}
{G}\left( S_{T}; {\cal M}, \tilde {\cal M}_{(1)},\ldots \right)  \rightleftharpoons  {\cal M}, \\
\tilde {G} _{(1)}( \ldots ) \rightleftharpoons  \tilde {\cal M}_{(1)},\,\tilde {G}_{(2)} ( \ldots ) \rightleftharpoons  \tilde {\cal M}_{(2)}\, \ldots  
\end{array} \right.
\label{GS}
\end{eqnarray}  

\begin{figure*}
	\begin{center}
		\includegraphics*[angle=270,width=155mm]{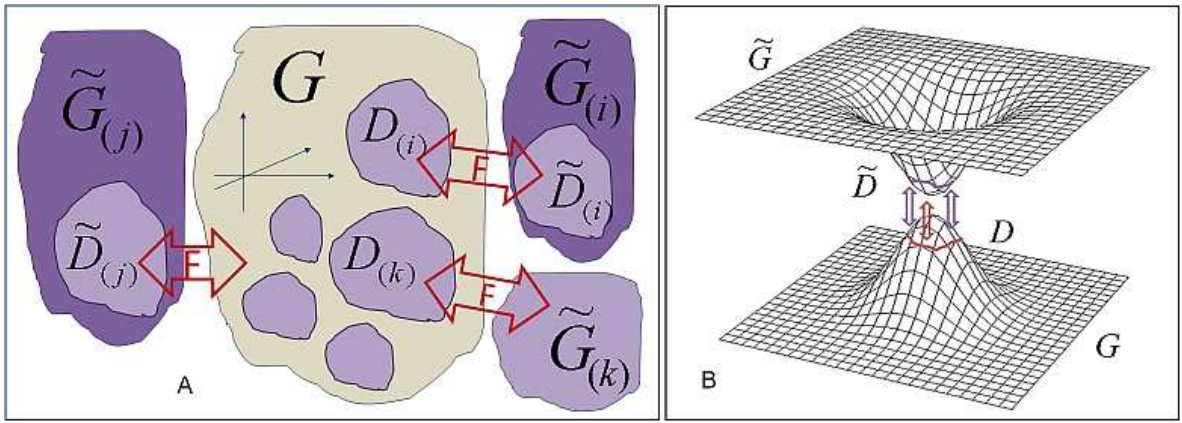}
	\caption{Off-site interaction between continuums} 
	\label{fig1}
	\end{center}
\end{figure*}

In the general case, not all matter, but some parts of it from \emph{action} regions $D\subset G$ and $\tilde D \subset \tilde G$ may be interconnected with each other. Three possible cases of interconnections between continuums are shown on Figure~\ref{fig1}A: the general case, when $G\supset D \rightleftarrows \tilde D_{(i)} \subset \tilde G_{(i)}$; and two particular cases, when $G \equiv D \rightleftarrows \tilde D_{(j)} \subset \tilde G_{(j)}$ and $G\supset D_{(k)} \rightleftarrows \tilde D_{(k)} \equiv \tilde G_{(k)}$. As far as the off-site matter takes part in gravitational interaction in the observer's continuum, it bends the observer's space-time. It is schematically shown on Figure~\ref{fig1}B. Thus, we have introduced the principle of \emph{off-site action} between continuums:
\begin{itemize} 
\item \emph{Continuums are in \emph{off-site action}, if corresponding parts of matter from their action regions are involved in gravitational interaction in each of them. } 
\end{itemize}

Note running ahead, that the notion of off-site continuums cannot be reduced to model of continuum with additional hidden or compact spaces. We will consider this question in details in next Section~\ref{sec:PoC}.  

In spite of the supposition of the existence of off-site matters, their existence meets principal objections. The first one: \emph{``if we have no particular answer to a question: ``what physical situation does this describe?'', then it remains an interesting case, but perhaps only in the mathematical sense''} (L.~Ryder, 2010). Other objections come from the theory of cognition, but they occur in a close connection with the first one.

\section{\bf Paradigm of Cognoscibility}
\label{sec:PoC}

The paradigm of cognoscibility in science declares that the world around is cognizable. One usually interprets it as follows: the world around is accessible for human cognition and even if some unknown regions exist, they are only undiscovered yet. Thus, any `off-site matter' is considered as the ordinary one, but not discovered yet. Such interpretation has logical justification: if something can influence us, so we are able to perceive it and to cognize, if not, in practice, it is undistinguished for us from nonexistent. That is why science deals with cognizable real world, called \emph{physical reality}, and considers `off-site matters' and `outside worlds' as some speculations.

However, what does `cognizable' mean? The fundamental properties of objects in quantum physics, such as dualism, uncertainties, are declared as `outside the human imagination', `outside the everyday experience', etc. The quantum mechanics is considered as \emph{``an anti-intuitive discipline \ldots full of mysteries and paradoxes, which we do not understand well enough, but are able to use''} (M.~Gell-Mann Ref.\cite{GM81}, 1981). Moreover, the observations in modern cosmology lead to the \emph{``absurd model''} of the universe, where \emph{``most of the universe is made of something fundamentally different from the ordinary matter we are made of. \ldots The contribution of ordinary matter to the overall mass-energy budget has been shown to be small, with more than 95\% of the universe existing in new and unidentified forms of matter and energy''} (W.~Freedman, M.~Turner Ref.\cite{Turn03}, 2003). We may also point out to `practically discovered' black holes, which explicitly do not belong to the space-time of the observer, so off-site and unreachable for him. All of this casts doubts on the above interpretation of the paradigm of cognoscibility and needs us to analyze in more details the process of human cognition. 

The observer creates the system of representations about the world around by interpreting events and interconnecting them by cause-effect chains. Such approach gives a possibility to predict events and to survive in this world. According to the existing paradigm of cognoscibility, we believe, that it is possible to interconnect by cause-effect chains all events at any infinitesimal interval in the space-time and to arrange them. It predetermines the use of real numbers for the space-time description, predetermines the continuality and connectivity of the space-time for us, i.e. predetermines the representation of the space-time as a continuum. Indeed, the affine connectivity and metrics are usually introduced for the space-time description in relativistic theories. Thus, the existing representations about the world around in physics are the consequence of using the causality principle in cognition. 

However, why are we so sure, that it is possible to interconnect by cause-effect chains \emph{all} events? It is known from the mathematical theory of sets (see, for example, Refs.\cite{Can874, Jech99}) that continuums are only one type of possible sets. They are classified as having the transfinite cardinal number $\aleph_1$, but, for example, countable sets have less cardinality $\aleph_0$ and the real functions of \emph{any} kind from real argument have higher cardinality $\aleph_2$. There also exist other sets with cardinalities $\aleph_3, \aleph_4 \ldots$ up to infinity. From this point of view it is not clear why we put in correspondence to the world around the space-time continuums, but not other sets? It seems that the only reason for us to use continuums is a convenience to arrange events by means of cause-effect chains with the logical `background' `proved experimentally' that it gives us a possibility to survive. Probably, it does not look like enough reason for objective representations\ldots. 

Let's consider few simple examples with sets to make the situation more clear. Let's $\{ n \}$ and $\{ m \}$ are two sets of integer numbers. Sets of integer numbers are `closed' in relation to operations of summation, subtraction and multiplication, because $\forall n_i,n_j \in \{n\}$: $ n_i + n_j, n_i - n_j, n_i n_j \in \{n\}$. It is possible to say that sets of integer numbers form the \emph{compact group} to these operations. If $n=m$ sets are equivalent to each other. However, if we `shift' them from each other onto some value $a$, i.e. putting $n=m+a$ (or $m=n-a$), sets will coincide when $a$ is integer and will not even intersect with each other otherwise, because in the last case $\forall m$: $m + a \notin \{n\}$ and $\forall n$: $n - a \notin \{m\}$, so sets $\{ n \}$ and $\{ m \}$ seem `parallel' to each other in relation to above operations. Which set $\{ n \}$ or $\{ m \}$ is now a set of integer numbers? -- it depends on the `observer', i.e. on choice of the `initial' frame for him. Of course, any set is appropriate to be `initial'. If one set, say for definiteness, $\{m\}$ is shifting in `time' $t$ with constant velocity $v$ from the set $\{n\}$, so $a=vt$, the observer in $\{n\}$, who is able to operate only with integer numbers, will interpret the set $\{m\}$ as `usual' or `ordinary' one only in discrete points of the real axis and as ``something fundamentally different'' in others. 

One can continue. When we `shift' from each other sets of rational numbers $\{ p \}$ and $\{ q \}$, putting $p=q+a$ (or $q=p-a$), so if $a$ is some irrational number, sets also do not `intersect' and may be considered as `parallel' to each other in relation to above operations plus division, i.e. sets may be considered as compact groups to these operations. One can approximate any point on real axis with any desired accuracy by rational numbers, so, at first sight, it may seem enough for the approximation of real numbers by rational ones. However, rational numbers form a countable set with the cardinality $\aleph_0$, while the cardinality of real numbers is $\aleph_1$, so irrational numbers are much more powerful than rational ones. It means, for example, that the possibility to observe the rational numbers on the real axis is equal to zero and it is impossible to put in point-to-point correspondence the rational and irrational numbers. Rational numbers may always be considered only as approximation to irrational ones, one may come from rational to irrational number only in passage to the limit. If, again as in previous example, one set, say for definiteness, $\{q\}$ is shifting in `time' $t$ with constant velocity $v$ from the set $\{p\}$, so $a=vt$, the observer in $\{p\}$, who is able to operate only with rational numbers, will interpret the set $\{q\}$ as `usual' or `ordinary' one only in countable points of the real axis and as `something fundamentally different' in others. Of course, the `amount' or cardinality of `fundamentally different' is incomparably larger than `ordinary'. 

If we analogously consider two sets of real numbers $\{x\}$ and $\{y\}$, i.e. \emph{continuums} with the cardinality $\aleph_1$, and shift them from each other, putting $x=y+\chi$ (or $y=x-\chi$), where $\chi$ is an element of the set with higher cardinality, say $\aleph_2$, so the observer in $\{x\}$ will interpret the set $\{y\}$ as the `ordinary' one only when $\chi$ is real and as `fundamentally different' in other cases. Again one will be able to reach the set of $\aleph_2$ from $\aleph_1$ only in passage through the limit and the `power' or cardinality of `fundamentally different' will be incomparably larger than `ordinary'. In the general case, it seems impossible to arrange elements in sets with the cardinalities higher than $\aleph_1$. We have mentioned in previous Section~\ref{sec:ST} that the notion of off-site continuums cannot be reduced to model of continuum with additional hidden or compact spaces, because continuums of any dimensionality have the transfinite cardinality $\aleph_1$, and not more. By the way, any $1D$-segment of real axis has the transfinite cardinality $\aleph_1$, so is as \emph{powerful} as any multidimensional continuum. 

Basing on this preliminary analysis, we may assert the following: 
\begin{itemize}
\item The representations of the world around created with help of causality correspond to a set of elements or events \emph{continuously} interconnected with each other. Mathematically it is a continuum with the transfinite cardinality not exceeded $\aleph_1$. 
\item There are no objective reasons to limit the representations of the world around by the continuum, so the space-time continuum may be considered only as some approximation of the world around. 
\item The space-time representations form a closed logical system for the internal observer, i.e. the compact group in relation to his cause-effect chains, so spaces exceeded the space-time continuum are unreachable for the observer's causality, and, in this sense, are `off-site' for him. 
\end{itemize}

The closure of the observer's space-time is a crucial point, but not the end of cognition, because this group is compact only for the observer's causality. The off-site phenomena would become `reachable' when we take in consideration interconnections and sets with higher transfinite cardinalities $\aleph_2, \aleph_3 \ldots$. That's why off-site continuums in Global system on Figure~\ref{fig1} were shown as some `compacts' or closed systems with mutual interconnections exceeded the cause-effect chains inside each continuum. 

We may specify the new interpretation of Paradigm of Cognoscibility as: 
\begin{itemize}
\item {\emph{The world around is cognizable, but its cognition with help of causality is limited and not complete}}. 
\end{itemize}

Now we have a background to consider the concept of off-site phenomena with hope to interpret mentioned above `mysteries and paradoxes' in quantum physics and `fundamentally different' substances in modern cosmology. 

\section{\bf Off-Site Action}
\label{sec:OSA}

In relativistic theories any \emph{identified} physical object may be described by the Action:  
\begin{equation}
S = \int^{t_2}_{t_1} dt L (q_j, \dot q_{j}) = \int^{t_2}_{t_1} dt \int {\cal L}(q_j,\partial_i q_{j}) \sqrt{-g} dV,  
\label{S}
\end{equation}
where $g$ is the determinant of metric tensor $g_{ik}$ in the observer's continuum, the action region $D$ is separated onto time $\tau = ct$ and space parts $D$: $\tau \times V$, $dx^0dx^1dx^2dx^3 \equiv d\tau dV$, $c$ is the speed of light. The Lagrangian $ L $ and the Lagrangian density $ {\cal L} $ are functions of some quantities $q$ and their derivatives $\partial_i q\equiv \partial q /\partial x^i$, $\dot q \equiv c\partial_0 q \equiv \partial_t q$ characterizing the state of the physical system. It defines the energy-momentum tensor $T_{ik}$ of physical object, which in GR is a part of the Einstein field equations: 
\begin{equation}
R_{ik}-\frac{1}{2}g_{ik}R=\frac{8\pi {\cal G}}{c^4} T_{ik}, 
\label{EFE}
\end{equation}
where $R= g^{ik}R_{ik}$ is a curvature, $R_{ik}$ a Ricci tensor, $\mathcal{G}$ a gravitational constant. 
 
At first sight, Eq~(\ref{S}) describes any physical object, so the ordinary and off-site matters are simply undistinguished for the observer. However, a careful examination reveals a principal difference between ordinary and off-site matters. E.~Schr\"odinger Ref.\cite{Schr50} had written about the Einstein field equations (see Eq~(\ref{EFE})) that: \emph{``I would prefer you to consider these equations not as field equations, but as a \emph{definition} of the energy-momentum tensor $T_{ik}$. As well as the Laplace's equation $\mathrm{div} E = \rho$ (or $\nabla^2 \phi =-4\pi \rho$) tells us nothing more than the divergence $E$ differs from zero everywhere where there is a charge, and we call $\mathrm{div} E$ as a charge density. The charge does not \emph{force} the electric vector to have the non-zero divergence, it exactly \emph{is} this non-zero divergence. Just the same, the matter does not \emph{force} the geometric value from the left side of this equation to differ from zero, it \emph{is} this non-zero tensor, it is described by {him}''}. 

One can see that matter in GR is identified with its influence onto the observer's continuum without considering the matter itself. The material point is quite appropriate model for such approximation, because the region of the matter existence is negligibly small and one can neglect its content replacing it by some macroparameters of the matter's influence. It defines the representation of the point mass by generalized functions $m(x_0)=m \int \delta(x-x_0) dx$, where $\delta$ is a delta-function.  The material point is a mathematical trick to take the fundamental question of the `origin' of matter out of consideration, or, at some level of cognition, a brilliant trick to avoid this consideration. By default, the `ordinary matter' is understood in GR as a \emph{distribution} of material points or, more generaly, `point objects' (vectors, tensors, etc.), so the ordinary matter is not a matter itself, but some characteristics of its influence onto the observer's space-time. Each material point, each part of ordinary distributions need to obey the physical laws in the observer's continuum, so scalar, vector, tensor fields are usually considered for ordinary matter description. 

In GS representations the off-site matter interacts with the observer's continuum on some action region $D\subset G$ (see Figure~\ref{fig1}). The observer cannot perceive the off-site matter directly, but outside the action regions $D$ the influence of off-site matter needs to coincide with the Action $S$ Eq~(\ref{S}), so the fields $\chi$ initiated by the off-site matter in $D$ need to satisfy to the \emph{Action integral}: 
\begin{equation}
\tilde S = \int^{t_2}_{t_1} dt L (q_j, \dot q_{j}) = \int^{t_2}_{t_1} dt \int_{V} {\cal L}(\chi) \sqrt{-g} dx,  
\label{AI}
\end{equation}
which differs from Eq~(\ref{S}) exactly by taking into account the action regions. Generally speaking, the initiated fields by themselves depend on the off-site matter and, in distinguish to ordinary distributions of material points, do not need to be interconnected by cause-effect chains in the observer's continuum, so do not need to obey the physical laws in it. Note, that we would not be able to introduce the initiated fields without reconsideration in Section~\ref{sec:PoC} the paradigm of cognoscibility in science. 

We can immediately say that as far as initiated fields $\chi$ and the action region $D$ depend on the off-site matter, they cannot depend explicitly on the off-site space-time continuum, so the action integral needs to be invariant. For $L$ to be invariant, the correlation between ${\cal L}$ and $D$ needs to exist. We will show in Section~\ref{sec:Q} that these correlations lead to quantization, but now we will consider for simplicity that ${\cal L}$ and $D$ are invariant both. Thus, $L$ is a scalar and ${\cal L}$ a scalar field. This way, quantities $q$ from Eq~(\ref{S}) characterizing the physical object are, in fact, macro-characteristics of initiated fields, so Eqs~(\ref{S}) and (\ref{AI}) have clear interconnection. For example, the point mass $m(x_0)=m \int \delta(x-x_0) dx$ may be considered as the passage through the limit with $\| V \| \rightarrow 0$ of the action integral Eq~(\ref{AI}) .  

The initiated fields seem not defined well enough inside the action region by Eq~(\ref{AI}), so different models and interpretations are possible. Below we will consider some of them, but we need to note before, that Eq~(\ref{AI}) defines on $D$ some functional space. Generally, these functions do not described by the observer's causality, so are of \emph{any} kind and, as it is known from the set theory, have cardinality $\aleph_2$. We will interpret models and in `understandable' notions of analytical functions of $\aleph_1$, but they will be only some approximations to more powerful multiformity.   

The requirement of ${\cal L}$ to be a scalar field on $D$ may be satisfied in different ways. In tensor algebra one can represent the scalar field as follows: 
\begin{eqnarray}
{\cal L} = \mu + A^i B_{i} + C^{ik}D_{ik} + E^{ijk}G_{ijk} + \cdots. 
\label{RR}
\end{eqnarray}  

In relativistic theories the terms of ${\cal L}$ are understood as the `ordinary' distributions of matter in the observer's space-time (so $D\equiv G$). The first term ${\cal L}=\mu$ has sense of the gravitating matter distribution or the mass density. The vector field $B_i$ may be associated with currents of electric charges $B_i =j_i$, so $A^i$ will be a $4$-vector potential and $C^{ik}=F^{ik}$, $D_{ik}=F_{ik}$ $4$-tensors of the electromagnetic field. 
 
The notion of initiated fields gives additional principal possibilities. If some terms of ${\cal L}$ (or their components) are considered as initiated fields, so others are \emph{complemented} ones to them to fulfill ${\cal L}$ to the scalar field. For example, it will be shown in Section~\ref{sec:Q} that considering currents of electric charges $B_i =j_i$ as initiated fields in action region $D\subset G$, the complemented fields $A^i$ and $F^{ik}$ will be extended to electroweak interactions. 

Using the notion of the initiated field one can get even more considerable generalization, because it occurs that the action integral Eq~(\ref{AI}) can include both relativistic and quantum approaches. Indeed, now, one can represent some terms of ${\cal L}$ as a scalar complex field $\psi(x)$, because, in distinguish to ordinary fields, it does not contradict to the notion of the initiated fields, because the complex conjugated field $\overline\psi(x)$ may successfully fulfill ${\cal L}$ to a scalar field: ${\cal L}=\overline\psi(x)\psi(x)$. One may continue these representations for spinor or matrix complex fields as follows:
\begin{eqnarray} 
{\cal L} = \mu+ \overline\psi \psi + \overline \eta^i \eta_i+ \overline \xi^{ij} \xi_{ij}+ \cdots. 
\label{QR}
\end{eqnarray} 
Here the summation over repeating indices is meant and up indices mean also a transposition, so $\eta^i$ is a component of a row-vector, $\eta_i$ of a column-one, and $\xi_{ik}=\xi^{ki}$. 

In fact, the representation Eq~(\ref{QR}) forms on $D$ the infinite dimensional complex vector space. It exactly coincides with the mathematical apparatus of quantum mechanics. Indeed, the eigenfunctions of the Hermit operators on this vector space form a full basis on $D$ and corresponding real-valued eigenvalues are quite appropriate for the description of macro-parameters `as a whole' of off-site objects. Objects `as a whole' need to satisfy to the physical laws in the observer's continuum, so, according to the well-known algorithm of the quantum mechanics, putting the eigenvalues $q_k$ of the Hermit operator $L_k=-\imath\partial_k$ in a correspondence to the components of the energy-momentum $4$-vector $p^k=(\mathcal{E}/c,\bm p)$ of a `particle' with the mass $m$ and applying $\hbar q_k = p_k$ in $p^k p_k =m^2 c^2$, one can get the Klein-Gordon equation $-\hbar^2 \partial^k\partial_k \psi = m^2c^2\psi$. The Schr\"odinger equation $-\imath\hbar c \partial_0 \psi = (\hbar^2 /2m) \partial^\alpha \partial_\alpha \psi + U(x^\alpha)\psi$ comes from $\mathcal{E}'=\bm {p}^2/2m + U(x^\alpha)$, where $\mathcal{E}'$ and $U$ are the kinetic energy and the potential. The Dirac equation for the spinor $\Psi=(\psi_0,\ldots,\psi_3)$ (the third term in Eq~(\ref{QR})) is a result of matrix factorization of the Klein-Gordon equation. For generality and discussion we may propose the model with the interconnection of off-site matters from two or more continuums on the same action region: $D\rightleftharpoons \tilde D_{(i)} \rightleftharpoons \tilde D_{(j)} \cdots$, which may be responsible for the description of the Quantum ChromoDynamics objects. 

The concept of off-site phenomena has important consequences. The quantum and relativistic theories occur the particular cases of GS representations, so many postulates and paradoxes of quantum physics have got clear interpretations. For example, the particles' duality, the wave function collapse, the EPR-like effects do not look like paradoxes in GS, because the initiated fields are not the `ordinary' distributions, so are not defined by cause-effect chains in the observer's continuum. The observer can identify the initiated fields with ordinary matter only `as a whole', in the material point approximation, but not in general. Correspondences and differences between relativistic and quantum approaches become quite obvious in GS: the physical objects are off-site for the observer's space-time, so off-site for his causality. Therefore, relativistic and quantum theories have different causalities, and their backgrounds are and need to be inconsistent with each other. 

The observation and the possibility of identification become in a very close connection with each other. In fact, in relativistic theories the observer interprets the world around from his space-time, from his `humancentric' point of view. The observer himself defines the privileged position of GR. \emph{``General relativity is, conceptually, a completely different sort of theory from the other field theories, because of its explicitly geometric nature''} (L.~Ryder Ref.\cite{Ryd09}), but `multicentric' GS representations can include quite consistently them both. 

\section{\bf Quantization}
\label{sec:Q}

The correlation through the initiated fields between relativistic Eq~(\ref{RR}) and quantum Eq~(\ref{QR}) representations gives interesting physical interpretations. For example, using the notion of initiated fields the electromagnetic fields may be extended up to electroweak interactions and stable in time off-site objects are quantized, which occurs the consequence of the energy conservation law in the observer continuum. 

In relativistic theories the physical objects with electric charges and electromagnetic fields are described in Eq~(\ref{S}) by the Lagrangian density ${\cal L}= \mu+ A^i j_i+ F^{ik} F_{ik}$, where $j_i$ are currents of electric charges, $A^i$ a $4$-vector potential and $F^{ik}$ $4$-tensors of electromagnetic fields. The variation of ${\cal L}$ leads to well-known wave equations for electromagnetic fields. If currents $j_i$ are understood as initiated fields, one needs to consider the action integral Eq~(\ref{AI}) instead of Eq~(\ref{S}). But the action integral is defined on the action regions $D$: $\tau\times V$, so for the wave equations the edge conditions $A^i, F^{ik}=0$ in $G\setminus D$ need to be satisfied: 
\begin{eqnarray} 
\partial^k \partial_k A^i=  j^i;\quad j^i, A^i \in D.  
\label{WE}
\end{eqnarray} 
Here we neglect the gravitational influence of off-site matter to the observer's continuum, because it is negligibly small on the scale of elementary particles in comparison to others, i.e. electroweak and strong. It is considered $\sqrt{-g}=1$ on $D$ for simplicity, but without loosing the generality. 

The electromagnetic fields cannot exist outside the action region $D$, because otherwise the stable in time off-site object would be an infinite energy source (or a drain), so edge conditions $A^i, F^{ik}=0$ in $G\setminus D$ are responsible for the energy conservation in the observer's continuum with exception of the stationary electric and magnetic fields, which do not transfer the energy. These edge conditions may be satisfied by the \emph{mechanism of compensation} when off-site sources are synchronized and compensate the electromagnetic fields excited by them outside the action regions. Thus, sources $j^i$ need to be self-matched with electromagnetic fields $A^i$ excited by them.
\begin{figure*}[ts]
	\begin{center}
		\includegraphics*[angle=270,width=155mm]{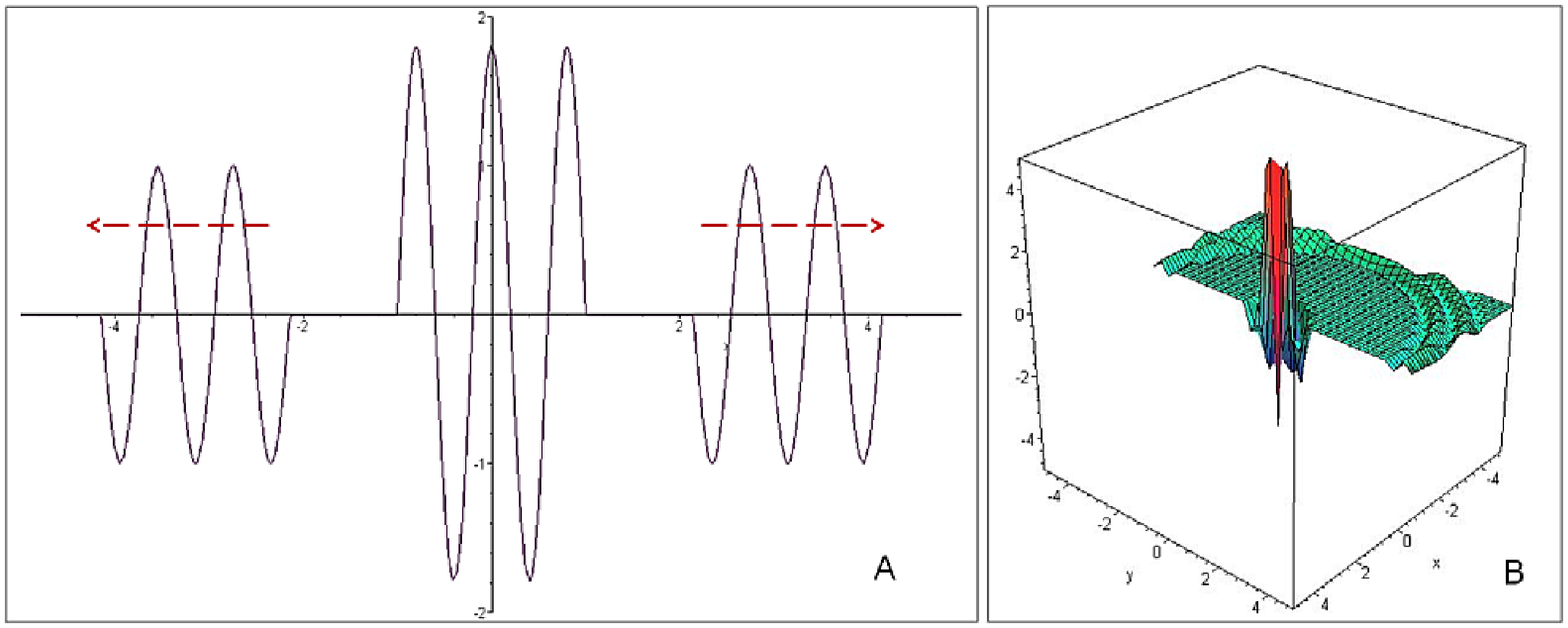}
		\caption{Excited fields: 1D (A) and 3D (B) cases.} 
	\label{fig2}
	\end{center}
\end{figure*}

Solving Eq~(\ref{WE}) in complex amplitudes $j(\tau, \bm r)= {\cal J} (\bm r) e^{\imath \omega^{(l)} \tau}$, where $\omega^{(l)} $ is the frequency and $x=(\tau,\bm r)$, one may come for the excited fields $u^{(l)} =U(\bm r)e^{\imath \omega^{(l)} \tau}$ to the stationary Helmholtz equation $\left( \nabla^2 + |\bm{k}^{(l)} |^2 \right)U(\bm {r}) ={\cal J} (\bm {r} )$, $k^2= |\bm {k}^{(l)}|^2=(\omega^{(l)})^2$, which in spherical polar coordinates $\bm r = (r,\vartheta, \varphi)$ has the fundamental solutions for sources ${\cal J} =\delta(r-R)Y^m_l(\vartheta,\varphi) $, where $Y^m_l$ are spherical Bessel functions (Ref.\cite{Vlad81}): 
$$
u^{(l)}_{jm}= e^{-\imath \omega^{(l)}_{j} \tau}\frac{C}{\sqrt{r}} J_{l+\frac{1}{2}}\left( k^{(l)}_{j} \frac{r}{R}\right) Y^m_l(\vartheta,\varphi). 
$$
Here $J_{l+\frac{1}{2}}$ are Bessel functions, $C$ is some constant. The edge conditions lead to the quantization equations $J_{l+\frac{1}{2}}(k^{(l)}_{j}R)=0$ defining the quantized spatial sizes of physical objects. 

It is convenient to describe the mechanism of compensation in $1D$ case (see Figure~\ref{fig2}A). Here two synchronized sources in points $p$ and $q$ excite waves on $x$-axis. If the \emph{quantization conditions} $q-p=\lambda(n+\frac{1}{2})$, $\lambda={2\pi}/{k}$, $n\!=\!0,\!1,\!\ldots$ are satisfied, there are the standing waves $U=\frac{1}{k}\sin k(x-p)$ on $x\in [p,q]$, while $U=0$ on `external' regions $x\in (-\infty,p)\cup(q,+\infty)$. With the allowed discrete spatial `sizes' two permanent sources become `self-matched' with each other and do not emit waves, so may be stable in time. Thus, only discrete states $S_n$ with spatial sizes $q-p=\lambda (n+\frac{1}{2})$ are allowed and the lowest state $S_0$ corresponds to $q-p=\lambda /2$. Standing waves have an energy, so the passing between states $S_n$ and $S_m$ are accompanied by emission (or absorption) in opposite directions of two field quanta with `length' $\lambda (n-m)$ if $n>m$ ($n<m$). These moving quanta are shown on Figure~\ref{fig2}A with arrows. The corresponding standing waves in spherically symmetric $3D$-case are presented on Figure~\ref{fig2}B. 

The observer interprets the standing waves as if some sources are in permanent exchange by field quanta with each other, exactly as it is represented in field theories. The electromagnetic fields excited by initiated ones seem `captured' in action regions, which has clear correspondences with the electroweak interactions. However, excited fields are not a reason of `confinement', they just complement the initiated fields defined on action regions. We will consider the `capturing' of the initiated fields in more details in next Section~\ref{sec:OSO}. Fields complemented to the initiated fields of higher terms in Eqs~(\ref{RR},\ref{QR}) will possess other conserved values with corresponding conservation laws. Taking into account that any particle may annihilate with its antiparticle, considered mechanism of compensation may be generalized for higher order fields, which needs to have correspondences with interactions in Quantum ChromoDynamics. 

It is possible to observe the self-matched compact electromagnetic fields in macro-objects, in so-called optical solitons (see, for example, N.~Rosanov, Ref.\cite{Ros07}, 2007) existing in non-linear optical media. The nonlinearities of optical media play a role of initiated fields of off-site sources. Certainly, optical solitons differ from the off-site objects by nature, but the way of formation of compact electromagnetic fields looks quite similar. Solitons also possess quite explicit quantum characteristics. There are fundamental and vortical solitons, which can interact and merge with each other, can create dynamically stable systems or divide to different parts. The fireballs also seem as compact electromagnetic fields created by self-matched sources in excited media, for example, in the ionized air. 

The correlation supposed in Section~\ref{sec:OSA} Eq~(\ref{AI}) between ${\cal L}$ and $D$ defines the quantization of off-site objects. There are no limitations on the spatial `sizes' of off-site objects in Eq~(\ref{WE}), but the complemented fields of `larger' objects may be destroyed by more energetic smaller ones, so the larger objects will be unstable because of these external influences. Thus, the stable off-site objects are seen by the observer with quantized spatial `sizes' and their spatial `scale' is determined by the energy level of the `surroundings'. Quantum postulates and paradoxes acquire quite clear interpretations with the concept of off-site phenomena. 

\section{\bf Off-Site Observations}
\label{sec:OSO}

The physical analysis of off-site phenomena in cosmology differs considerably from previous analysis. In previous sections we have investigated the \emph{stable} in time and \emph{identified} physical objects. Generally, it is not inherent for large-scale cosmological objects, so we have for analysis only the notion of the initiated fields and GS interpretations given by the set theory. We will start from obvious mathematical examples and, after that, will come to physical interpretations and descriptions different from the `mainstream' one. 

The off-site objects may be observed \emph{externally}, outside the action region, i.e. from regions $G\setminus D$, or \emph{internally} inside the action region $D$. Exactly the action regions are out of consideration in relativistic theories, approximated by `point properties' in them and other continuums do not exist at all. Thus, the situations of internal observations seem quite paradoxical in GR.  

Mathematically, many GS interpretations are based on the set theory, where the subset may be as powerful as the set it belongs to. For example, the transformations $x^i=\arctan \tilde x^i$ between coordinates of two identical $n$-dimensional continuums $G$: $\{ (x^0{,}\ldots{,}x^{n-1}) \}$ and $\tilde G$: $\{ (\tilde x^0{,}\ldots{,}\tilde x^{n-1}) \}$ map `point-to-point' the continuum $\tilde G$ into the $n$-dimensional cube (a subset $D$) of  continuum $G$, so $\tilde G\rightleftharpoons D \subset G$. On the other hand, with the reversed transformations $\tilde x^i=\arctan x^i$ the continuum $G$ is mapped into $\tilde G$, so $G \rightleftharpoons \tilde D\subset \tilde G$.  
\begin{figure*}[ts]
	\begin{center}
		\includegraphics*[angle=270,width=155mm]{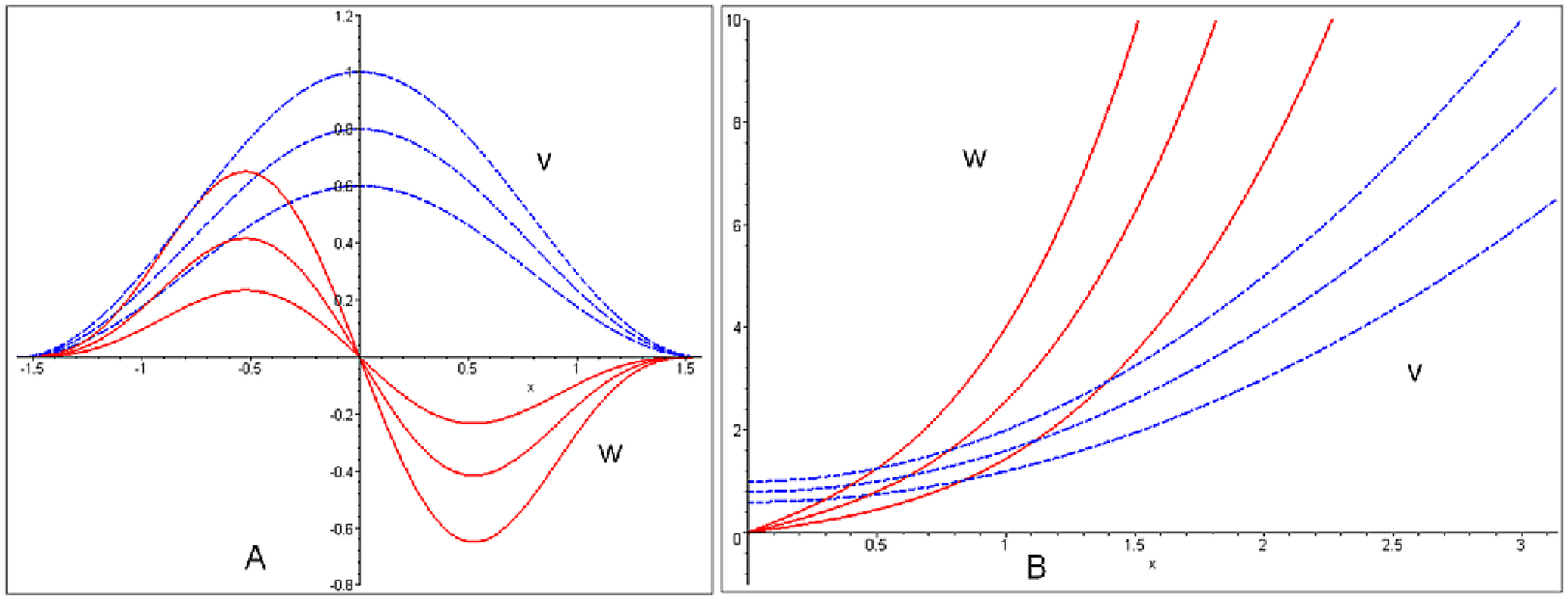}
	\caption{The external (A) and internal (B) observations of `off-site test particle' (the dash line - the velocity, the solid line - the acceleration).}
	\label{fig3}
	\end{center}
\end{figure*}

Thus, the limitation of off-site continuum in action regions of the observer's continuum does not mean that off-site continuum is limited by itself. The free test particle from $\tilde G: (\tilde \tau, \tilde r)$ moving along the axis $\tilde r$ with the constant velocity $\tilde v$ (so, its acceleration $\tilde w = 0$) will be seen in $G: (\tau, r)$ connected with $\tilde G$ by coordinate transformations $r=\arctan \tilde r$ and $\tau=\tilde \tau$ as moving with the velocity $v=\tilde v \cos^2 r$ and the acceleration $w=\tilde v^2 \sin 2r \cos^2 r$. The test particle seems `confined' inside $(-\pi/2,\pi/2)$, because its visual velocity $v \rightarrow 0$ with $r \rightarrow \pm \pi/2$, and the acceleration seems centripetal (see Figure~\ref{fig3}A). The considered transformations are not the only ones. For example, in case of $\tilde r= - \frac{r}{|r|} \ln (1-|r|)$,  $r\in (-1,1) \leftrightarrows (-\infty, \infty) \ni \tilde r$ one will have: $ v=\tilde v (1-|r|) \stackrel {r \rightarrow \pm 1}{\longrightarrow} 0$ and $w=-{\tilde v}^2 \frac{r}{|r|} (1 -|r|)$, so off-site matter seems as captured in interval $(-1,1)$. Such `capturing' is off-site and so unexplainable for the observer in GR, he needs to introduce new fundamental for him interaction - the strong one to determine a `reason'. 

Let's consider the last examples of `confinement' in case of internal observation, i.e. when the observer's continuum $G$ is mapped into the $n$-dimensional cube (a subset $D$) of the off-site continuum $\tilde G$, so $G \rightleftharpoons \tilde D\subset \tilde G$. It corresponds to transformations reversed to just considered, i.e. $\tilde r = \arctan r$ and $r\in (-\infty,\infty)\leftrightarrows \tilde r \in (-\pi/2,\pi/2)$. Now the free test particle from off-site continuum will be seen in the continuum of the observer as moving with the velocity $v = \tilde v (1+r^2) \rightarrow \infty$ with $r \rightarrow \pm\infty$ and with the centrifugal acceleration $w=2\tilde v^2 r (1+r^2)$ (see Figure~\ref{fig3}B). Thus, during internal observations the off-site objects seem as `flying away' from each other with acceleration. The accelerated expansion of the Universe is now observed in cosmology. Transformations $\tilde r= \frac{r}{|r|}(1-e^{-|r|})$, $r\in(-\infty,\infty)$: $v=\tilde v e^{|r|} \stackrel {r \rightarrow \pm \infty}{\longrightarrow} +\infty$, $w= {\tilde v}^2 \frac{r}{|r|}e^{2|r|}$ lead to exponential expansion like in the de Sitter model of the universe. 

Considered mathematical examples illustrate the principal possibilities that the set theory may give for physical description. In the general case, deterministic transformations cannot describe the off-site action, but we used them to make connections with existing causal representations, to make principally new models obvious and understandable for \emph{our causal kind of thinking}. 

For these reasons, we will investigate the `internal structure' inside the action regions by making analogies with the curvilinear frames of references. It is possible to consider any curvilinear frame $\tilde S$ in GR, but not all of them may be considered as \emph{`physically realized'} by `real bodies' (L.~Landau, E.~Lifschitz Ref.\cite{LL88}), only if the metric tensor $\tilde g_{ik}$ of $\tilde S$ satisfies to: 
\begin{eqnarray} 
\begin{array}{l}\tilde g_{00} >0,\quad \tilde g <0,\\ \left|\begin{array}{ll}\tilde g_{00} &\tilde g_{01} \\ \tilde g_{10} &\tilde g_{11} \end{array}\right| <0,\end{array} \left|\begin{array}{lll} \tilde g_{00} &\tilde g_{01} &\tilde g_{02} \\ \tilde g_{10} &\tilde g_{11} &\tilde g_{12}\\ \tilde g_{20} &\tilde g_{21} &\tilde g_{22} \end{array}\right| >0 .
\label{Cond} 
\end{eqnarray} 
These conditions guarantee the time $d\tilde\tau=\sqrt{\tilde g_{00}}dx^0$ and length $d\tilde l^2=(- \tilde g_{\alpha \beta} + \tilde g_{0 \alpha} \tilde g_{0 \beta} /\tilde g_{00}) dx^\alpha dx^\beta$, $\alpha,\!\beta\! = \! 1{..}3$ in curvilinear frame to have the physical sense. The initiated fields may not satisfy to conditions Eq~(\ref{Cond}), because they exactly exceed `real bodies' (in fact, this notion coincides with `ordinary matter') and exactly `unphysical objects' are now observed in cosmology. 

To analyse the internal structure of off-site objects, we will conditionally separate the action regions on areas, where the element $\tilde g_{00}$ and determinant $\tilde g \equiv \det \tilde g_{ik}$ of off-site metric tensor are seen as positive or negative (see Ref.\cite{NB08}) in the observer's continuum. 

The \emph{timelike area} $(\tilde g_{00}>0$ and $\tilde g<0)$: The visible off-site time $d\tilde\tau$ and length $d\tilde l$ have physical sense, so the observer may identify the off-site objects as the `ordinary matter', but their motion \emph{inside action regions} is defined \emph{mainly} (see below for details) by topology of their continuums, so seems unusual in the observer's continuum as if acted by some `invisible forces': a confinement or, vise versa, accelerated expansion on Figures~\ref{fig3}A,B correspondingly.  

The \emph{transitive area} $(\tilde g_{00}<0$ and $\tilde g<0)$: \emph{``The non-fulfillment of the condition $\tilde g_{00}>0$ would mean only, that the corresponding frame of references cannot be realized by real bodies; thus if the condition on principal values is carried out, it is possible to achieve $\tilde g_{00}$ to become positive by appropriate transformation of coordinates''} Ref.\cite{LL88}. The space-time parameters of off-site objects cannot be identified by the observer and these objects will be interpreted as something amorphous, distributed in space, but it is possible to identify their influence to other objects, because they can `act like real bodies'. For example, in rotational frame of references the real massive body behind the horizon of events cannot be identified by the rotational observer, but the gravitational influence of this `off-site' body to other bodies inside the horizon wouldn't disappear. The off-site objects from transitive areas correlate with `dark matter' defined in cosmology as \emph{``some invisible distributed in space substance, strange `amorphous' media interacting gravitationally with identified visible objects''}. 

The \emph{spacelike area} $(\det \tilde g_{ik}>0)$: \emph{``The tensor $\tilde g_{ik}$ cannot correspond to any real gravitational field at all, i.e. the metrics of the real space-time''} Ref.\cite{LL88}. The space-time parameters of off-site objects also cannot be identified by the observer and objects will be interpreted as something amorphous, distributed in space, but, in distinguish to transitive area, there are no possibility to identify their action with `real bodies', so off-site objects will possess quite unusual, even unphysical characteristics (such as, for example, the gravitational repulsion or negative pressure). In cosmology such unusual characteristics are inherent to `dark energy'. 

We have emphasized before that the visual properties of off-site objects are determined \emph{mainly} by off-site `topology', because due to our postulate of off-site action in Section~\ref{sec:ST} both matters from observer's and off-site continuums are involved in interaction. Thus, the dynamics of ordinary and off-site matter depend on them both. It means that instead of the invariance of the general integral $I=\int {\cal R} dx$ of the observer's continuum in GR, one needs to consider the invariance of the `global integral' $I+\sum \tilde I_{(j)}$ of global system Eq~(\ref{GS}). So, generally, one needs to consider the variation:
\begin{equation}
\delta \left(I + \sum \tilde I_{(j)}\right) = 0. 
\label{I}
\end{equation}
Now the variation $\delta I = 0$ considered in GR is only a particular case of Eq~(\ref{I}) when off-site objects are stable, so $\sum \tilde I_{(j)} = 0$. Exactly this condition is we have used for analysis of quantum physical objects in previous sections. Generally, $ \delta I = - \sum \delta \tilde I_{(j)} \neq 0$, which defines the `matter exchange' between continuums, so there is not even the energy conservation in the observer's continuum for nonstable off-site objects -- there is the `hidden' for the observer the energy conservation in the global system instead. 

Formally, Eq~(\ref{I}) leads to Einstein field equations Eq~(\ref{EFE}) with additional `initiated' energy-momentum tensor $T^*_{ik}= \sum_j T^{(j)}_{ik} $ (different action regions may intersect, so, this way, we are supposing the superposition of off-site actions) taking into account the initiated fields of off-site objects: 
\begin{equation}
R_{ik}-\frac{1}{2}g_{ik}R = \frac{8\pi {\cal G}}{c^4} \left( T_{ik} + T^*_{ik} \right). 
\label{GFE}
\end{equation}
From one side, coming from Eq~(\ref{I}) to Eq~(\ref{GFE}) we may lose the solutions, because $T^*_{ik}$ need to be tensor fields, which are not so powerful as the initiated ones, but, from other side, this approximation seems as the only available way for the observer to perceive off-site objects, so  Eq~(\ref{GFE}) is convenient for interpretation with using the above separation on areas inside action regions. 

For example, let's consider in non-relativistic case the off-site macroscopic continuous bodies described in their own continuum by the energy-momentum tensor $\tilde T^{ik}$ with non-zero elements $\tilde T^{00}=\varepsilon$ and $\tilde T^{\alpha \alpha}=p$, $\alpha\!=\!1{..}3$, where $\varepsilon$ is an energy density and $p$ the pressure. These off-site objects will be seen in the observer's continuum in metrics defined by the off-site metric tensor $\tilde g_{ik}$. In space-like areas of the action regions one needs to have $\det \tilde g >0$, so it is possible to consider $\tilde g_{ik}= -1$ if $i\!=\!k$ and $\tilde g_{ik}= 0$ if $i\!\neq\!k$. This way $\det \tilde g = 1>0$, and one can get for $T^* = \tilde g_{ik} \tilde T^{ik} = - \varepsilon - 3 p$. The usual condition for $T^* >0$ leads to: $\varepsilon < - 3 p$ and with $\varepsilon > 0$ to $p<0$, so describes the substance with negative pressure. When $G \rightleftharpoons \tilde D\subset \tilde G$ this substance seems distributed in whole continuum of the observer and has clear correspondences with dark energy observed now in cosmology with the approximated equation of state $p = -\varepsilon$. Note, that `dark energy' in GR is only a model to interpret the accelerated expansion of the universe and some other gravitational effects. The models of `dark substances' do not come from `standard representations' of GR, moreover, seems quite `unusual' and `unphysical' for them, while in GS they have quite clear interpretation as an off-site observations. 

There are clear physical correspondences of off-site phenomena with objects of quantum physics and objects beyond `the ordinary matter we are made of' in cosmology, which are just different perceptions or approximate representations of off-site matter. 

\section{\bf Levels of Cognition}
\label{sec:LoC}

In the concept of off-site phenomena the space-time is frames convenient for the observer and even built by him to arrange events in the world around with help of cause-effect chains. It is our method of cognition, our neurophysiological possibilities. \emph{``All kinds of individual reactions are now related with the spatial organization of a brain, the character of associations of neurons in micro- and macro-ensembles, their arrangement, the relations with each other and with other ensembles''} (see Refs.\cite{Adr83, Bl85}). 

Amounts of neurons limits the possibility of human brain to reflect events of the world around. From the set theory point of view, it may be estimated as having a cardinality $\aleph_0$ of a discrete countable sets. We will define such reflection as a zero level of cognition. The using of the causality principle is defined by the neurophysiology of our brain\footnote{ The detailed analysis of the process of cognition by the human brain shows that the `image' of the world around is `constructed' as the system of notions with help of mechanism: ``thesis-antithesis-synthesis''. At first, neurons in brain continuously excited from sensory organs are interconnected (generalized) by interneuronic synapses, axons. It is a `thesis'. Then, this `model' from `short-time' memory is compared again with signals from sensory organs with help of the hippocamp, tonsils and hypothalamus of the brain (`antithesis') and is corrected until the coincidence (`synthesis'). Only after that the brain saves the `final model' in long-time memory (see Ref.\cite{Bl85}). }. It lets to extend the possibilities of cognition to sets of continuums and continuous functions with the cardinality $\aleph_1$. Certainly, such extension leads to corresponding losses of contents in `models' (notions) and their interconnections during `generalization'. This is the first level of cognition, the level of logical systems, deterministic methods in mathematics, classical and relativistic theories. The existing paradigm of cognoscibility asserts that the world around may be described completely on this level of cognition. 

Staying on this level, the observer tries to interconnect by cause-effect chains all available phenomena in the world around, declaring other phenomena, even if they exist, as not important and `undistinguished from nonexistent', so the observer tries to create the rigorous logical theory by default. General relativity seems the most general theory on this level of cognition. The GR has \emph{``background independent nature''} Ref.\cite{Smol06}, because its basic notions are included in the system, so the system seems logically `closed': we are searching the events which may be included in the logical system and are able to perceive them only on frames of existing logical system. The space-time in GR is exactly the framework of our representations, in which we are trying to consider all events of the world around. The closure of logical system related to cause-effect chains warrants the introduction of off-site systems, off-site space-times. It is quite difficult to realize the closure of own logical system, our space-time and to overcome its causal limits. In fact, the Global system declines the `humancentric' (in analogy with heliocentric) model of the world around by declaring the possibility of existence of different points of view, different logically separated causalities, geometries, worlds. Probably, such `multi-verses' are different sides of Something single with universal interconnections exceeded our causal understanding. 
\begin{itemize}
\item {\emph{ The world around is that we perceive, but he is not obliged to be exactly like this}}. 
\end{itemize}

\begin{figure}[ts]
	\begin{center}
		\includegraphics*[angle=270,width=77mm]{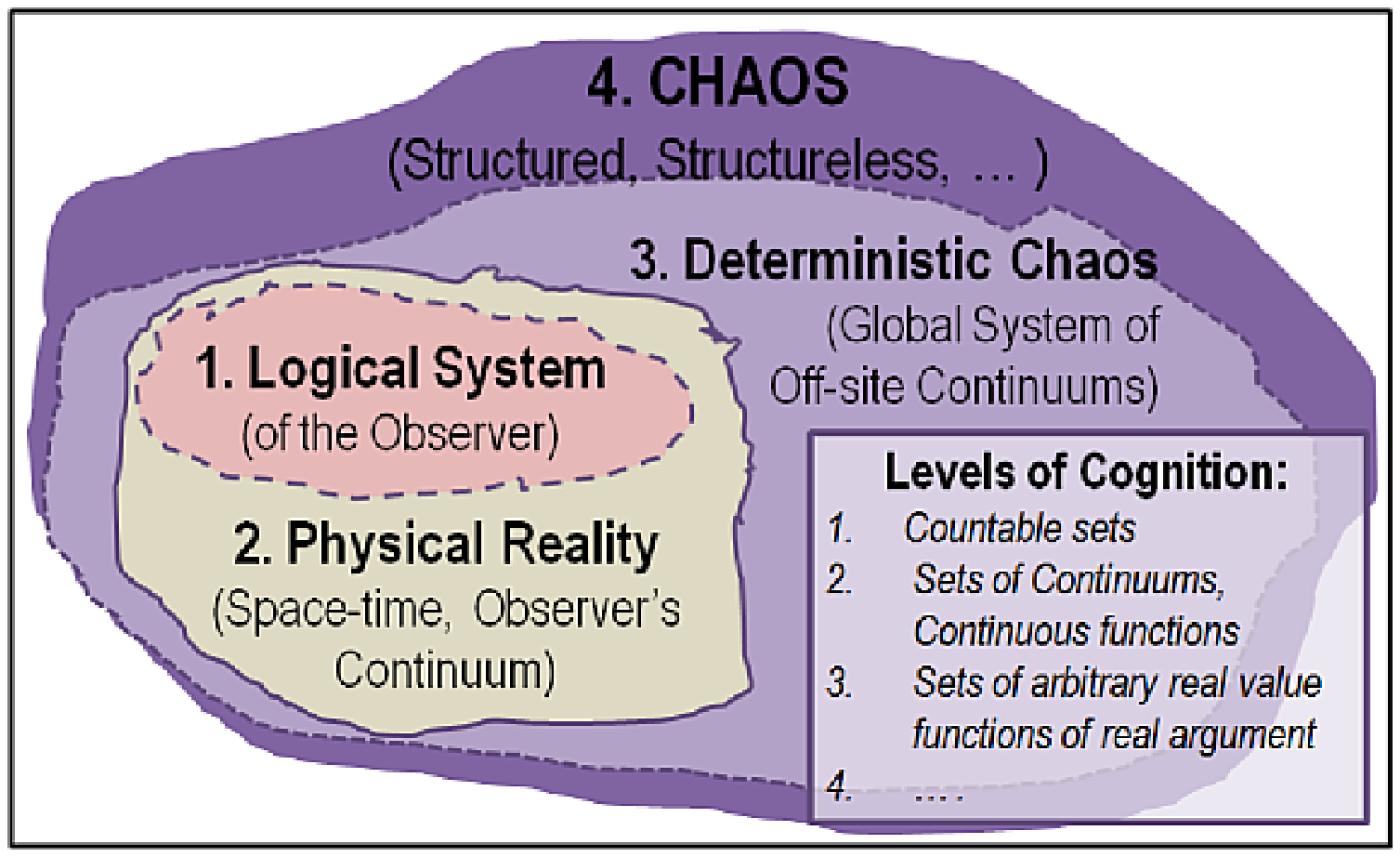}
	\caption{Levels of Cognition.}
	\label{fig4}
	\end{center}
\end{figure}

At a moment, there are no reasons to limit the `power' of the world around by some level corresponding to sets with some fixed transfinite cardinality. According to the analysis in Section~\ref{sec:PoC}, we may also suppose by analogy that each level of cognition may be `logically closed' in relation to some available operations with its elements. Taking into account the existence of sets with high cardinalities up to infinity, one may suppose the multi-level structure of the world around from the cognition point of view (see Figure~\ref{fig4}). The world around seems `opened', because there are the infinite number of levels of cognition, while each level may seem `logically closed'. 

We have used some statements of the set theory to describe the levels of human cognition, but it seems that the concept of off-site phenomena, in particular connected with the theory and with the neurophysiology of human cognition, may, in its turn, become useful for the set theory, which seems not installed well enough and is still under construction. 

Experiments in quantum physics and cosmological observations give quite explicit evidences of existence of objects and phenomena in the world around principally exceeded the GR model, off-site for our causality. The world around seems much richer, more powerful, than our causal representations about it. The world look chaotic to us, when we are going to overcome our causal frames and it needs to be to. The manner of description of off-site phenomena cannot look so `clear and quantitative' as deterministic methods inside logical system and they need to be like this, other possibilities need to be found. For example, nobody can predict the exact place of registration of the electron on the screen during interference or diffraction, and it is a principal statement, but one can predict quite precisely the electrons' probability distribution. Thus, such methods as probabilistic description, fractals, generalized functions, $p$-adic models, extended reals, etc. (see also E.E.~Rosinger Ref.\cite{Ros09} and references in it) may be considered as efforts to operate with sets of high cardinalities. 

Using the concept of off-site phenomena different levels of cognition and methods of description on each level may be separated from each other. Thus, classical and relativistic theories operates in observer's space-time, so are connected with the observer and seems logical systems for him, while the objects of quantum physics and dark substances in modern cosmology are off-site for the observer, for his logical system. It does not mean that off-site objects are something different from `ordinary matter'. The ordinary matter is only an approximated model or logical perception of much more powerful matter in the world around. According to such `unification' of relativistic and quantum theories in global system, it is obvious that their `standard unification' in frames of the rigorous logical theory seems quite doubtful. 

The world around seems more chaotic and disordered than we think before. These representations surprisingly correlate with epistemic views of ancient philosopher Plato: \emph{`the man has the only possibility to see the distorted shadows of the bright and multicolor Reality on the curved wall of the cave, where he is confined and chained up back to the entrance'}. We are similarly `captured' by cause-effect chains inside our logic and can only try to reconstruct the `multicolor Reality' by its images reflected on our `curved logical systems', basing on the `distorted shadows' given by our senses. This is a way of human cognition, but, it seems, not a limit. 

\section*{\bf Conclusion}

The physical reality is that we perceive with help of causality, but the world around seems much more complicated. The space-time continuums differed from the observer's one need to exist. The paradigm of cognoscibility has got corrected interpretation: the world is cognizable, but the cognition of it with help of causality is limited and there are many levels of cognition depending on the considered degree of chaos. The relativistic and quantum theories operate on different levels of cognition, so their `unification' seems doubtful. We have no reasons to interrupt the cognition of the world around on some fixed level, so a number of levels of cognition may be infinite.

\end{document}